# Quantum Yield Calculations for Strongly Absorbing Chromophores

Stephen P. Nighswander-Rempel

Centre for Biophotonics and Laser Science
School of Physical Sciences, University of Queensland
Brisbane, QLD  Australia 4072
Email: snighrem@physics.uq.edu.au

**Summary:** This article demonstrates that a commonly-made assumption in quantum yield calculations may produce errors of up to 25% in extreme cases and can be corrected by a simple modification to the analysis.

The radiative quantum yield is an important quantity in molecular chemistry. Defined as the fraction of molecules (of a particular compound) that emit a photon upon direct excitation by a source,[1] it is a measure of the proportion of de-excitation that occurs radiatively. Quantum yields provide important information regarding excited electronic states, radiationless transitions, and coupling of electronic to vibronic states. Moreover, they are used in the determination of chemical structures, sample purity, and appropriateness of laser media.[2-4] For these reasons, it is important to have reliable methods of calculating accurate quantum yield values for new substances.

Whilst the oldest and most fundamental methods for calculating the quantum yield are based on Vavilov's method of measuring absolute luminescences,[5] these methods are difficult and require great precision. They have established certain compounds such as quinine sulphate and anthracene as standards with well-accepted quantum yield values,[6, 7] but it is now more common to calculate quantum yields of new compounds by comparing emission rates to those of a known standard. These studies assume that the quantum yield is proportional to the ratio of fluorescence emission integrated across the spectrum to the absorption coefficient at the excitation wavelength.[8-10] However, it has been recently shown that the quantum yield is actually proportional to the ratio of integrated emission to $(1-e^{-\alpha d})$, where $\alpha$ is the absorption coefficient and $d$ is the width of the excitation volume from which the detected fluorescence is emitted.[11] When the absorption coefficient and the sample thickness are small, these methods are roughly equivalent. However, for substances with large coefficients (e.g. the natural pigment melanin, Fig. 1), this assumption is invalid.

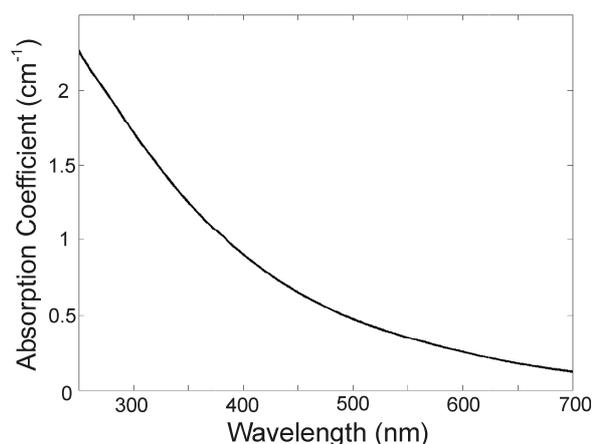

Figure 1. Absorption coefficient spectrum for 0.005% synthetic eumelanin.

It should be noted that this is not merely a result of the inner filter effect, whose implications have been discussed previously,[1, 6, 12-14] but rather stems directly from the definition of the quantum yield. Although the quantum yield is defined as the ratio of emitted photons to absorbed photons,[1, 12] this is not equivalent to the ratio of emission intensity to absorbance.[11] Even when the inner filter effect is corrected for, use of this approximation in the calculation of quantum yield is still invalid. In this communication, the error incurred by this approximation is reported, showing that it is more significant than other errors which are currently taken into account in precise calculations of quantum yield. Synthetic eumelanin is used as an example due to its strong absorbance.

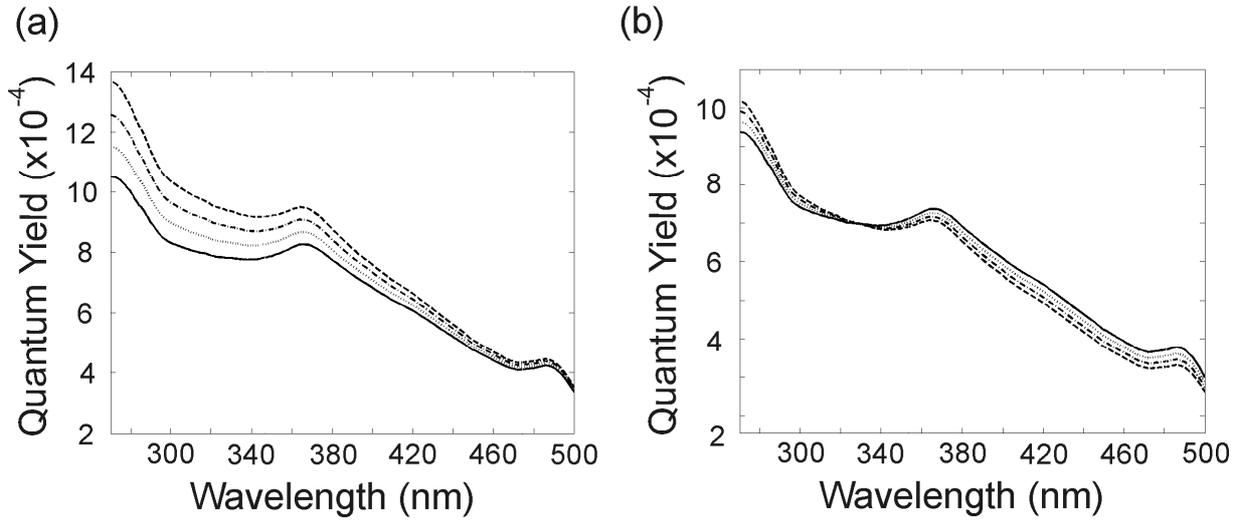

Figure 2. (a) Dependence of quantum yield of melanin on excitation wavelength. Quantum yield values are assumed to be proportional to either the ratio of integrated emission to absorbance (solid line) or to the ratio of integrated emission to $(1-e^{-\alpha d})$, for $d$ values of 0.1 cm (dot), 0.2 cm (dot-dash), or 0.3 cm (dash). (b) Quantum yield values calculated using higher quinine sulphate concentrations ($1 \times 10^{-6}$ to $1 \times 10^{-4}$ M) for the reference standard.

Methods of solution preparation (synthetic eumelanin and quinine sulphate), absorbance and fluorescence spectroscopy, and calculation of the quantum yield were identical to those described previously.[11] In short, the quantum yield $\phi$ was shown to be

$$\phi(\lambda_{ex}) = \frac{\int I_d^*(\lambda_{ex},\lambda_{em})d\lambda_{em}}{(1-e^{-\alpha(\lambda_{ex})d_{ex}})} \frac{\phi_{st}(1-e^{-\alpha_{st}(\lambda_{ex})d_{ex}})}{\int I_{d,st}^*(\lambda_{ex},\lambda_{em})d\lambda_{em}} \frac{n_{sample}^2}{n_{st}^2} \quad (1)$$

where $I_d^*$ is the detected fluorescence intensity (corrected for probe attenuation, the inner filter effect, and variation in the beam intensity with wavelength), $\lambda_{ex}$ and $\lambda_{em}$ are the excitation and emission wavelengths, respectively, $\alpha(\lambda_{ex})$ is the absorption coefficient of the sample at the excitation wavelength, $d_{ex}$ is the width of the excitation volume in the sample, $n$ is the refractive index, and the subscript $st$ refers to measurements on the quantum yield standard (quinine sulphate).

When the absorption coefficient of the sample is small, $\alpha d$ is much less than one and the function within the parentheses is roughly equal to $\alpha d$ (using a first-order Taylor approximation of the exponential), yielding

$$\phi(\lambda_{ex}) = \frac{\int I_d^*(\lambda_{ex},\lambda_{em})d\lambda_{em}}{\alpha(\lambda_{ex})} \frac{\phi_{st}\alpha_{st}(\lambda_{ex})}{\int I_{d,st}^*(\lambda_{ex},\lambda_{em})d\lambda_{em}} \frac{n_{sample}^2}{n_{st}^2} \quad (2)$$

(note that factors of $d_{ex}$ in the numerator and denominator have cancelled). This is the relation commonly used in quantum yield studies.[8-10] Whilst Eq. 1 is more exact than Eq. 2 in that it does not make use of the Taylor series approximation, it does require knowledge of the value of $d_{ex}$. Since this value represents the thickness of the volume in the cuvette from which fluorescence reaches the detector, it is not easily measured. Therefore, quantum yield values calculated using Eq. 1 have been obtained for three values of $d_{ex}$: 0.1, 0.2 and 0.3 cm.

Resulting quantum yield values using both Eqs. 1 and 2 are displayed in Fig. 2a. First, it is clear that for increasing values of $d_{ex}$, the difference between the values calculated with Eq. 1 and Eq. 2 escalates due to correspondingly poorer approximations in the Taylor polynomial. Thus, the accuracy of the calculated quantum yield value is very dependent on the excitation volume width. For example, the absorption coefficient of the eumelanin solution used here at 250 nm is 2.27 cm$^{-1}$. Given an estimated excitation volume width of 0.1 cm, the quantum yield value calculated with Eq. 2 is in error by 10%. For $d_{ex}$=0.3 cm, the error jumps significantly to 25%. At 400 nm excitation, where the absorption coefficient is much smaller, the corresponding errors in the quantum yield are 4% and 10%, respectively.

These data demonstrate that the error in using Eq. 2 is magnified significantly for samples with high absorption coefficients, but use of Eq. 2 can produce significant errors even for samples with low absorbance if concentrated reference standards are used. For the data presented in Fig 2a, several quinine solutions, ranging in concentration from $1 \times 10^{-6}$ to $1 \times 10^{-5}$ M in 1 N $H_2SO_4$ solution, were measured and a linear regression was applied to determine the optimal ratio for both Eqs. 1 and 2. When more concentrated quinine solutions ($1 \times 10^{-6}$ to $1 \times 10^{-4}$ M) were included in the regression, the quantum yield values did not converge with increasing excitation wavelength (decreasing melanin absorption coefficient, Fig. 2b). This was because the error in the Taylor approximation on the high-concentration quinine solutions limited the available accuracy in the quantum yield values even when the melanin absorption coefficient is low. For this melanin solution, even at an absorption coefficient of 0.5 cm$^{-1}$ (500 nm excitation) and $d_{ex}$=0.1 cm, the resulting error is 5%. When only high concentrations of quinine are used ($1 \times 10^{-5}$ to $1 \times 10^{-4}$ M), this error is magnified further. This is an important consideration since quinine solutions of even greater concentration have been used previously in a highly cited paper.[6]

A curious feature in Fig. 2b is the presence of an isosbestic point. The position of the isosbestic point depends on the reference standard concentrations used and represents the excitation wavelength where the absorption coefficient of melanin matches that of the reference standard, quinine. At this point, the $(1-e^{-\alpha d})$ factors in the numerator and denominator of Eq. 1 cancel each other out of the equation. Thus, variation of $d_{ex}$ has no effect on the calculated quantum yield.

In summary, the assumption often made – that quantum yield is proportional to the ratio of integrated emission to absorption coefficient – is not strictly correct. Depending on the sample absorption coefficient and excitation volume width, this approximation can produce errors of up to 25%. This is in addition to inner filter effects and stems from the definition of the quantum yield.

Moreover, even when sample absorption coefficient values are small, use of high reference standard concentrations can limit the accuracy of calculated values to 5-10%. In contrast, acceptable precision for quantum yield values is on the order of a few percent,[15] and even a small variation of 0.25%/°C in quantum yield values of quinine has been described as a "steep temperature dependence".[6] Relative to errors such as temperature-dependence of the quantum yield and differences in refractive index and excitation wavelength between the sample and the reference standard (which are all routinely accounted for in accurate quantum yield measurements), the errors incurred by this approximation are large indeed. Given that they are removed by a simple alteration to the analysis, future quantum yield calculations should use Eq. 1, which follows directly from the mathematical definition of the quantum yield.